# Anomalous Hall effect in ferrimagnetic metal $RMn_6Sn_6$ (R = Tb, Dy, Ho) with clean Mn kagome lattice

Lingling Gao[1,2], Shiwei Shen[1], Qi Wang[1,3], Wujun Shi[4,5], Yi Zhao[1], Changhua Li[1], Weizheng Cao[1], Cuiying Pei[1], Jun-Yi Ge[6], Gang Li[1,3], Jun Li[1,3], Yulin Chen[1,3,7], Shichao Yan[1,3], and Yanpeng Qi[1]*

[1]School of Physical Science and Technology, ShanghaiTech University, Shanghai 201210, China

[2]University of Chinese Academy of Science, Beijing 100049, China

[3]ShanghaiTech Laboratory for Topological Physics, ShanghaiTech University, Shanghai 201210, China

[4]Center for Transformative Science, ShanghaiTech University, Shanghai 201210, China

[5]Shanghai high repetition rate XFEL and extreme light facility (SHINE), ShanghaiTech University, Shanghai 201210, China

[6]Materials Genome Institute, Shanghai University, 200444 Shanghai, China

[7]Department of Physics, Clarendon Laboratory, University of Oxford, Parks Road, Oxford OX1 3PU, UK

**ABSTRACT**

Kagome lattice, made of corner-sharing triangles, provides an excellent platform for hosting exotic topological quantum states. Here we systematically studied the magnetic and transport properties of $RMn_6Sn_6$ (R = Tb, Dy, Ho) with clean Mn kagome lattice. All the compounds have a collinear ferrimagnetic structure with different easy axis at low temperature. The low-temperature magnetoresistance (MR) is positive and has no tendency to saturate below 7 T, while the MR gradually declines and becomes negative with the increasing temperature. A large intrinsic anomalous Hall conductivity about 250 $\Omega^{-1}cm^{-1}$, 40 $\Omega^{-1}cm^{-1}$, 95 $\Omega^{-1}cm^{-1}$ is observed for $TbMn_6Sn_6$, $DyMn_6Sn_6$, $HoMn_6Sn_6$, respectively. Our results imply that $RMn_6Sn_6$ system is an excellent platform to discover other intimately related topological or quantum phenomena and also tune the electronic and magnetic properties in future studies.

* Correspondence should be addressed to Y.Q. (qiyp@shanghaitech.edu.cn)

The kagome lattice, made of two-dimensional a network of corner-sharing triangles[1], has become one of the most fundamental models for unique quantum phenomena in condensed matter physics. Theoretical works indicate that kagome lattice yield an unusual electronic structure, including flat bands and Dirac points, which have been confirmed by experiments[2-5]. The kagome lattice with spin-orbit coupling (SOC) and out-of-plane ferromagnetism realizes the spinless Haldane model[6], leading to opened Chern gap at the Dirac points[7, 8]. A number of well-known kagome magnets with exotic phenomenon have been discovered, such as bilayer Fe kagome ferromagnetic $Fe_3Sn_2$[7, 9, 10], ferromagnetic Weyl semimetal $Co_3Sn_2S_2$[11-14], and non-collinear antiferromagnet $Mn_3Sn$[15, 16] . These systems exhibit a large intrinsic anomalous Hall effect (AHE) due to the Berry phase effects and negative magnetoresistance arises from chiral anomaly and flat bands with negative orbital magnetism and so on[12-14]. However, the kagome layer in these systems is not a "clean" one with tin atoms located at the centers of hexagons.

Recently, $RMn_6Sn_6$ (R = trivalent rare earth elements) kagome magnet family has attracted growing interest due to the pristine Mn kagome lattice. Owing to the chemical pressure from R atoms, the Sn atoms move away from Mn kagome layer. Large intrinsic anomalous Hall conductivity (AHC) has been observed in in-plane ferromagnetic $LiMn_6Sn_6$, in-plane ferrimagnetic $GdMn_6Sn_6$, and helical antiferromagnetic $YMn_6Sn_6$[17-20]. A large Chern gap in $TbMn_6Sn_6$ with out-of-plane ferrimagnetism was found by scanning tunnelling microscopy (STM) measurements[8]. In addition, $YMn_6Sn_6$ also shows a topological Hall effect near the room temperature due to double-fan spin structure with c-axis components when the field is applied in plane[19]. The collinear ferrimagnetic $RMn_6Sn_6$ (R = Tb, Dy, Ho) with different easy axis gives us a chance to tune electronic and magnetic states and explore larger AHE and other phenomena.

In this work, we systematically study the transport and magnetic properties of $RMn_6Sn_6$ (R = Tb, Dy, Ho) with pure kagome layers of Mn. The $RMn_6Sn_6$ (R = Tb, Dy, Ho) single crystal have been grown by the self-flux method, and they are collinear ferrimagnet with different easy axis. The value of magnetoresistance (MR) gradually declines and changes from positive to negative with the increasing temperature. Through the results of transport and magnetic, we find that $RMn_6Sn_6$ (R = Tb, Dy, Ho) all have large AHC originated from the intrinsic mechanism.

Single crystals of $RMn_6Sn_6$ (R = Tb, Dy, Ho) were synthesized by Sn-flux method[21] with a molar ratio R : Mn : Sn = 4.5 : 27 : 95.5. R (purity 99.95%), Mn (purity 99.9%) and Sn (purity 99.99%) grains were mixed and placed in an alumina crucible, and then sealed in a quartz ampoule under partial argon atmosphere. The quartz ampoule was heated in a furnace and kept at 1100 °C for about 24 h, then cooled down to 600 °C over about one week. Finally, the quartz ampoule was moved quickly into the centrifuge to sperate the excess Sn flux. X-ray diffraction pattern was carried out at room temperature by a Bruker D8 X-ray machine with Cu $K_α$ radiation ($λ$ = 0.15418 nm). Electrical transport and magnetization measurements were performed using Quantum Design PPMS-9T and MPMS3. The longitudinal and Hall electrical resistivity were measured using a five-probe method. In order to remove the longitudinal resistivity contribution due to voltage probe misalignment, we extract the pure Hall resistivity by the equation $\rho_{yx}(\mu_0 H) = [\rho(+\mu_0 H) - \rho(-\mu_0 H)]/2$. Correspondingly, the longitudinal resistivity component was obtained using $\rho(\mu_0 H) = [\rho(+\mu_0 H) + \rho(-\mu_0 H)]/2$.

The first principles calculations were performed based on density functional theory as implemented in Vienna *Ab initio* Simulation Package (VASP)[22] with the projector augmented wave potential[23, 24]. The exchange-correlation potential was formulated by generalized gradient approximation with Perdew-Burke-Ernzerhof (PBE) functional[25]. A Γ-center 8×8×8 *k* points grid was used for the first Brillouin zone sampling.

STM experiments were carried out with a Unisoku low-temperature scanning tunneling microscope at the base temperature of 4.3 K. The HoMn$_6$Sn$_6$ single crystals were cleaved at 77 K under ultrahigh vacuum, and then transferred into the STM head for measurements.

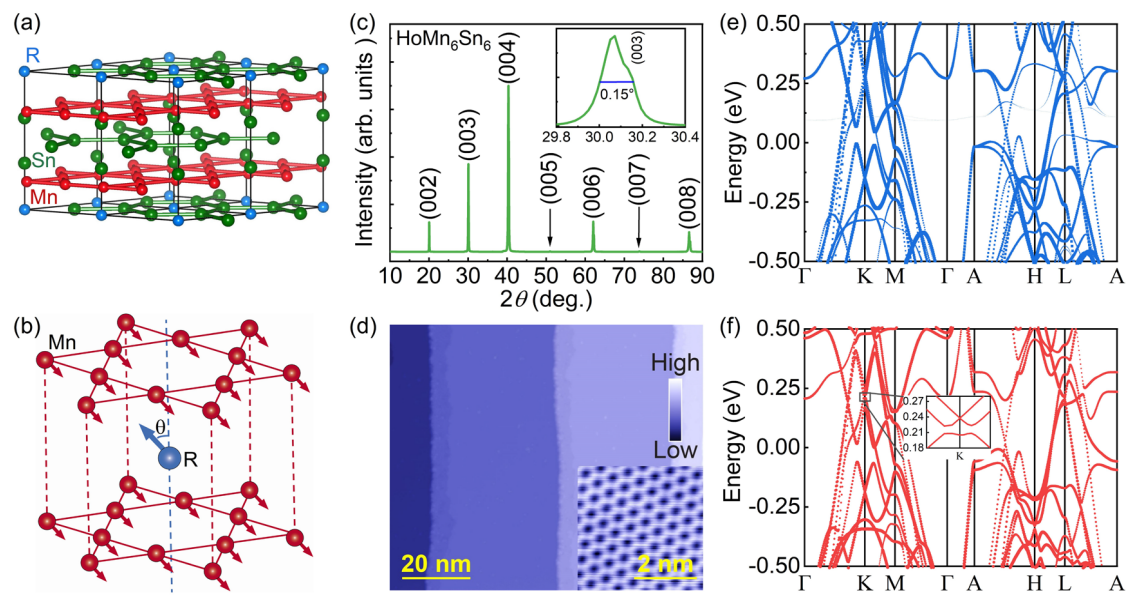

Fig.1. (a) Crystal structure of RMn$_6$Sn$_6$. (b) Illustration of magnetic structure of RMn$_6$Sn$_6$ at zero field at low temperature. (c) XRD pattern of HoMn$_6$Sn$_6$ single crystal with (00*l*) reflections at room temperature. The inset of the right panel zooms in the (003) reflection. (d) Constant-current STM topography on HoMn$_6$Sn$_6$ (V$_s$ = -200 mV, I = 50 pA). Inset is a zoom-in view of the typical cleaving surfaces (V$_s$ = -200 mV, I = 500 pA). (e) and (f) calculated band structure from Mn of HoMn$_6$Sn$_6$ without SOC and with SOC respectively. The size of dots represents the weight of Mn *d*-orbitals. The inset of (f) shows that Dirac points at *K* are gapped.

RMn$_6$Sn$_6$ (R = Tb, Dy, Ho) crystallizes in the hexagonal HfFe$_6$Ge$_6$-type structure with space group of *P6/mmm*[26]. As shown in Fig. 1(a), the crystal structure of all the compounds is built by stacking of RSn$_2$-Mn$_3$-Sn-Sn$_2$-Sn-Mn$_3$-RSn$_2$. In the RSn$_2$ layer, the Sn$_2$ atoms form graphene-like hexagonal lattice and the R atoms locate in the centers of the hexagons. In addition, the Mn$_3$ atoms form a kagome lattice but the Sn atoms located in the centers of the Mn hexagons is away from Mn$_3$ layer due to the chemical pressure of R atoms. Moreover, there is another Sn$_2$ layer forms a graphene-like hexagonal plane between Mn$_3$-Sn layers. The Mn-Mn sub-lattice is strongly ferromagnetic coupling and the R-Mn atoms are collinear antiferromagnetic coupling.

We define that the moment deviates from the [001] direction with $\theta$, and the $\theta$ for TbMn$_6$Sn$_6$, DyMn$_6$Sn$_6$ and HoMn$_6$Sn$_6$ are reported to be about 0°, 45° and 50° at low temperature, respectively (Fig. 1(b))[21, 27-29]. The XRD pattern of HoMn$_6$Sn$_6$ single crystal is exhibited in Fig. 1(c), the results of TbMn$_6$Sn$_6$ and DyMn$_6$Sn$_6$ are shown in Fig. S1 (Supporting Information). It reveals that the crystal surface is parallel to the *ab*-plane. The inset of Fig. 1(c) shows the full width at the half-maximum (FWHM) of the (003) peak is only 0.15. Fig. 1(d) shows typical cleaving surfaces of the HoMn$_6$Sn$_6$ measured by STM, indicating the high quality of single crystals. The band structures of HoMn$_6$Sn$_6$ from Mn sub-lattice calculated without and with SOC are shown in Fig. 1(e) and (f). The size of dots in Fig.1 (e) and (f) represents the weight of Mn orbitals. We observe linear band crossings at *K* point, above the Fermi energy at 215 meV. Considering the SOC, these linear crossings open a gap with 15 meV at *K* point (the inset of Fig. 1(f)). Similarly, the gaps at *K* and *K'* points are also observed in TbMn$_6$Sn$_6$ with 35 meV[8].

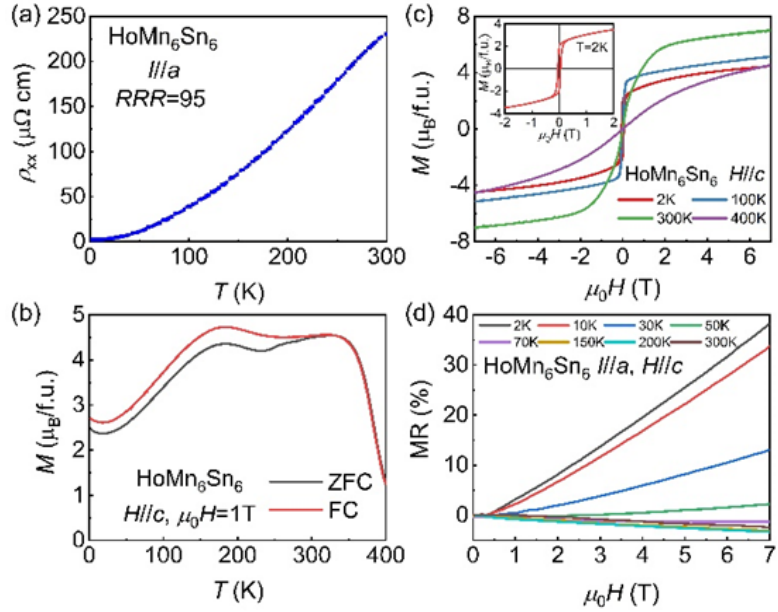

Fig.2. (a) Temperature dependence of the longitudinal electric resistivity $\rho_{xx}(T)$ for HoMn$_6$Sn$_6$. (b) Temperature dependence of magnetization $M(T)$ with ZFC and FC modes at $\mu_0 H$ = 1 T for *H//c*. (c) Field dependence of magnetization $M(\mu_0 H)$ with *H//c* at different temperatures for HoMn$_6$Sn$_6$. Inset: Magnetic hysteresis loop at 2 K with *H//c*. (d) Field dependence of magnetoresistance (MR) at different temperatures with *H//c*.

Fig. 2(a) exhibits the temperature dependence of longitudinal resistivity $\rho_{xx}(T)$ for the HoMn$_6$Sn$_6$ single crystal from 1.8 to 300 K. It shows a metallic behavior in the whole measuring temperature range. The residual resistivity ratio [RRR = $\rho_{xx}$ (300 K) / $\rho_{xx}$ (4.2 K)] is about 95, indicating the high quality of the sample. The $M(T)$ curves of HoMn$_6$Sn$_6$ with zero-field-cooling (ZFC) and field-cooling (FC) modes at $\mu_0 H$ = 1 T for *H//c* are shown in Fig. 2(b), $M(T)$ curves for TbMn$_6$Sn$_6$ and DyMn$_6$Sn$_6$ are shown in Figs. S2b and S3b respectively (Supporting Information). HoMn$_6$Sn$_6$ and DyMn$_6$Sn$_6$ present ferrimagnetic-type transition at $T_c \approx$ 383 and $T_c \approx$ 398 K respectively, which is determined by the peak of the derivative of magnetization $dM(T)/dT$. The Curie temperature of TbMn$_6$Sn$_6$ is about 423 K[21, 26-29]. Because it is beyond our measurement temperature range, we cannot determine $T_c$ here. The results are consistent with previous reports[21, 26-29]. At low temperatures, $M(T)$ curves exhibit an anomaly behavior at about $T_t = 200$ K, which could be related to spin reorientation from the *ab*-plane to the *c*-axis with decreasing temperature. It is consistent with the previous results of neutron diffraction[27, 28]. Due to the antiferromagnetic Mn-Ho coupling, the magnetic moment decreases with further lowering the temperature. The field dependence of magnetization $M(\mu_0H)$ at $T = 2$ K for $H//c$ shows a hysteresis, accompanying with a very low coercive field $\mu_0H_c \sim 0.05$ T (inset of Fig. 2(c)). When $T > T_c$, the hysteresis behavior vanishes and the $M(\mu_0H)$ curve presents a paramagnetic behavior (Fig. 2(c)). The data of magnetoresistance (MR = $[\rho_{xx}(\mu_0H) - \rho_{xx}(0)] / \rho_{xx}(0) \times 100\%$) are shown in Fig. 2(d). The sign of MR visibly changes at 50 K. When $T$ is below 50 K, the value of MR is positive and has no tendency to saturate below 7 T. However, when $T$ is above 50 K, the curves are negative. The positive MR at low temperature could attribute to the Lorenz force effects. On the other hand, the negative MR at high temperature could be related to the suppression of spin-disorder scattering. A similar evolution of MR and $M(\mu_0H)$ is also observed for TbMn$_6$Sn$_6$ and DyMn$_6$Sn$_6$ (Figs. S2, S3, Supporting Information), and the coercive field $\mu_0H_c$ at 2 K is about 0.8 and 0.35 T, respectively.

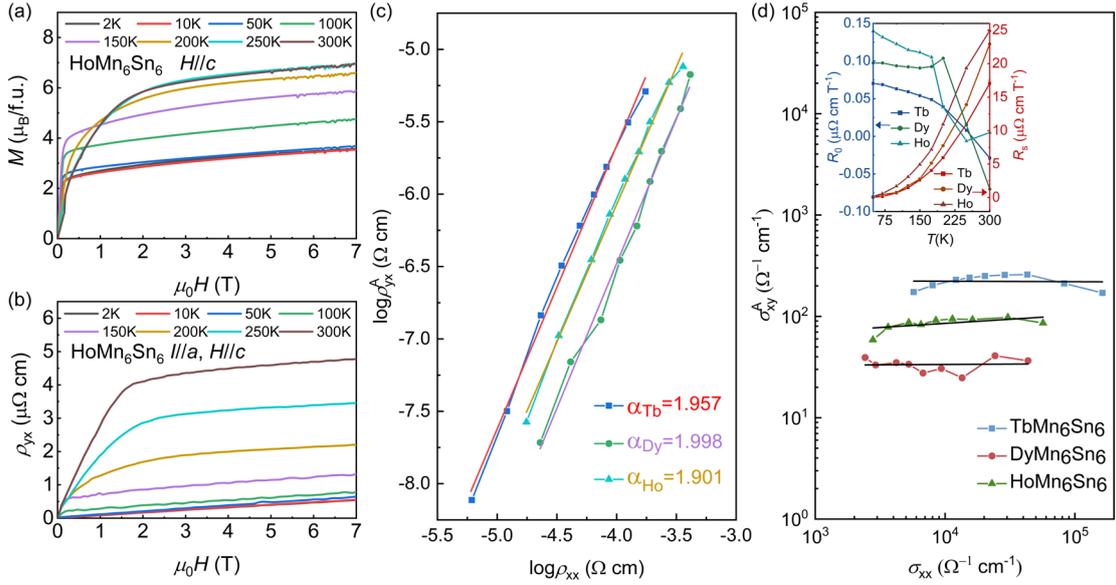

Fig.3. (a) Magnetization $M(\mu_0H)$ and (b) Hall resistivity $\rho_{yx}(\mu_0H)$ as a function of magnetic field from 0 to 7 T at different temperatures for HoMn$_6$Sn$_6$. (c) Plot of $log\rho_{yx}^A$ against $log\rho_{xx}$ for RMn$_6$Sn$_6$ (R = Tb, Dy, Ho). (d) The AHC $\sigma_{xy}^A$ plotted against longitudinal conductivity $\sigma_{xx}$ for RMn$_6$Sn$_6$ (R = Tb, Dy, Ho). The black solid lines represent the fitting line using the relation: $\sigma_{xy}^A \propto \sigma_{xx}^\beta$. The inset shows the temperature dependence of ordinary and anomalous Hall coefficients $R_0$ and $R_s$, respectively.

The field dependence of magnetization $M(\mu_0H)$ at various temperatures from 0 to 7 T with $H//c$ has been measured for HoMn$_6$Sn$_6$ (Figs. 3(a) and S5(a)). The shape of $M(\mu_0H)$ curves is typical for ferrimagnets, which increases rapidly at low field region with a saturation in the higher magnetic field. Figs. 3(b) and Fig. S5(b) exhibit the Hall resistivity $\rho_{yx}(\mu_0H)$ for HoMn$_6$Sn$_6$ as a function of magnetic field $\mu_0H$ at various temperatures. When increasing the temperature, the saturated value of $\rho_{yx}(\mu_0H)$ becomes bigger gradually and the $\rho_{yx}(\mu_0H)$ curves become weakly field-dependent at high field. However, at low temperature (below 50 K), the $\rho_{yx}(\mu_0H)$ curves show linear behavior. It indicates that the anomalous Hall resistivity becomes too weak to be observed, and $\rho_{yx}$ is dominant by the ordinary Hall effect contribution. The similar features are also observed in TbMn$_6$Sn$_6$ and DyMn$_6$Sn$_6$ (Figs. S4b and S4d) as well as other members of RMn$_6$Sn$_6$ family, such as GdMn$_6$Sn$_6$[17] and LiMn$_6$Sn$_6$[18]. Here, we mainly focus on the AHE above 50 K. It is conventional to express $\rho_{yx}(\mu_0H)$ in two parts[30],

$$\rho_{yx} = \rho_{yx}^0 + \rho_{yx}^A = R_0B + 4\pi R_sM \quad (1)$$

Where the first term $\rho_{yx}^0$ in the formula is the normal Hall resistivity and the second term $\rho_{yx}^A$ is the anomalous Hall resistivity. $R_0$ and $R_s$ represent the ordinary and anomalous Hall coefficient, respectively. Using equation (1), we are able to separate the contribution of $\rho_{yx}^0$ and $\rho_{yx}^A$ parts for RMn$_6$Sn$_6$.

The obtained $R_0$ is positive in a large temperature range for RMn$_6$Sn$_6$ (R = Tb, Dy, Ho) (the inset of Fig. 3(d)), indicating that the dominant carrier is hole-type. In addition, $R_s$ is positive in the whole temperature range and increases monotonically with increasing temperature, but the absolute value is much larger than $R_0$. The obtained $R_s$ value of HoMn$_6$Sn$_6$ is 24.93 μΩ cm T$^{-1}$ at 300 K, which is close to the value of Fe$_3$Sn$_2$[7, 9]. When fitting $log\rho_{yx}^A$ against the $log\rho_{xx}$ curve above 50 K with the formula $\rho_{yx}^A \propto \rho_{xx}^\alpha$, we find that the scaling exponent $\alpha$ is 1.957, 1.998, and 1.901 for HoMn$_6$Sn$_6$, TbMn$_6$Sn$_6$ and DyMn$_6$Sn$_6$, respectively (Fig. 3(c)). The nearly quadratic relationship clearly indicates that the AHE is not governed by the extrinsic skew-scattering mechanism in which $\rho_{yx}^A$ is linearly proportional to $\rho_{xx}$ [31-34]. By contrast, the intrinsic Karplus and Luttinger (KL)[32] or extrinsic side-jump mechanism[31] dominates the AHE in RMn$_6$Sn$_6$ (R = Tb, Dy, Ho). At temperature above 50 K, the AHC $\sigma_{xy}^A$ obtained by $\sigma_{xy}^A = \rho_{yx}^A / [(\rho_{yx}^A)^2 + \rho_{xx}^2]$, which is about 95, 250 and 40 Ω$^{-1}$cm$^{-1}$ for HoMn$_6$Sn$_6$, TbMn$_6$Sn$_6$ and DyMn$_6$Sn$_6$, respectively. The values of $\sigma_{xy}^A$ are comparable with those values in other RMn$_6$Sn$_6$ materials, such as YMn$_6$Sn$_6$ and GdMn$_6$Sn$_6$[17, 19, 20]. Theoretically, the intrinsic AHC $|\sigma_{xy,in}^A|$ is of the order of $e^2/(hc)$, where $e$, $h$, and $c$ is electronic charge, Plank constant, and cross-plane lattice constant, respectively[35, 36]. For HoMn$_6$Sn$_6$, the cross-plane lattice constant $c$ is 9.013 Å, corresponding to $|\sigma_{xy,in}^A| \approx 428$ Ω$^{-1}$cm$^{-1}$, which is comparable with the experimental value of $\sigma_{xy}^A$. On the other hand, the extrinsic side-jump contribution of $|\sigma_{xy,sj}^A|$ is expressed as $(e^2/hc)(\varepsilon_{SO}/E_F)$, where $\varepsilon_{SO}$ and $E_F$ are the spin-orbit interaction and Fermi energy, respectively[35, 36]. Generally, the $\varepsilon_{SO}/E_F$ is less than 0.01. Hence the $|\sigma_{xy,sj}^A|$ is much smaller than the obtained $\sigma_{xy}^A$ and the side-jump contribution should be negligible. In addition, the AHC $\sigma_{xy}^A$ shows a weak dependence on longitudinal conductivity $\sigma_{xx}$ (Fig. 3(d)). The absolute value of scaling index $\beta$ in $\sigma_{xy}^A \propto \sigma_{xx}^\beta$ equals 0.081, 0.005, and 0.008 for HoMn$_6$Sn$_6$, TbMn$_6$Sn$_6$ and DyMn$_6$Sn$_6$, respectively. Therefore, the AHE of RMn$_6$Sn$_6$ (R = Tb, Dy, Ho) at high field region is dominated by the intrinsic mechanism. The slightly various values of intrinsic AHC would be possibly tuned by the different R elements of RMn$_6$Sn$_6$.

In conclusion, we grow RMn$_6$Sn$_6$ (R = Tb, Dy, Ho) single crystal with clean Mn kagome lattice using the self-flux method and systematically study their transport and magnetic properties. A large intrinsic AHC of 250, 40, and 95 Ω$^{-1}$cm$^{-1}$ is observed for

TbMn$_6$Sn$_6$, DyMn$_6$Sn$_6$, and HoMn$_6$Sn$_6$, respectively. The results imply that RMn$_6$Sn$_6$ system is an excellent platform to study the relationship between the magnetic and electronic structure and helps to explore more kagome magnets with extraordinary properties.

See the supplementary material for x-ray diffraction pattern and, detailed magnetic and transport measurement of TbMn$_6$Sn$_6$ and DyMn$_6$Sn$_6$. It also contains the zoomed in plot of magnetization and transverse conductivity for HoMn$_6$Sn$_6$.

This work was supported by the National Key R&D Program of China (Grant No. 2018YFA0704300), the National Natural Science Foundation of China (Grant No. U1932217, 11974246, 12004252, 61771234), Natural Science Foundation of Shanghai (Grant No. 19ZR1477300), and the Science and Technology Commission of Shanghai Municipality (No. 19JC1413900). The authors thank the support from C$h$EM (No. 02161943) and Analytical Instrumentation Center (No. SPST-AIC10112914), SPST, ShanghaiTech University. W. S. acknowledges support from Shanghai-XFEL Beamline Project (SBP) Grant No. 31011505505885920161A2101001. The calculations were carried out at the HPC Platform of ShanghaiTech University Library and Information Services.

## DATA AVAILABILITY

The data that support the findings of this study are available from the corresponding authors upon reasonable request.